# Link Before You Share: Managing Privacy Policies through Blockchain

Agniva Banerjee
Dept. of Computer Science and Electrical Engineering
UMBC
agniv1@umbc.edu

Dr. Karuna Pande Joshi
Information Systems Department
UMBC
karuna.joshi@umbc.edu

*Abstract*—With the advent of numerous online content providers, utilities and applications, each with their own specific version of privacy policies and its associated overhead, it is becoming increasingly difficult for concerned users to manage and track the confidential information that they share with the providers. Users consent to providers to gather and share their Personally Identifiable Information (PII). We have developed a novel framework to automatically track details about how a user's PII data is stored, used and shared by the provider. We have integrated our Data Privacy ontology with the properties of blockchain, to develop an automated access-control and audit mechanism that enforces users' data privacy policies when sharing their data across third parties. We have also validated this framework by implementing a working system *LinkShare*. In this paper, we describe our framework on detail along with the LinkShare system. Our approach can be adopted by Big Data users to automatically apply their privacy policy on data operations and track the flow of that data across various stakeholders.

## I. INTRODUCTION

Providers of online content and services, including Big Data providers and e-commerce sites, often capture large amount of Personally Identifiable Information (PII) of their end users. This is primarily done to make their offering more user friendly and ensuring a seamless user experience [23]. This is also done by providers to determine usage patterns and to track their user community to be able to provide customized services which is critical to build customer loyalty [27]. Providers generate data privacy policy documents for their services and acquire consent from their end users on how their PII data will be acquired, stored, managed and used by the providers.

Data Privacy policies contain difficult to comprehend legalese in their terms and conditions, so end-users often do not pay attention to what they are agreeing to when they accept the privacy terms and conditions, and have little knowledge of how that data is used or shared by the service provider. Also, the data stored by service providers are often repurposed for carrying out research, without the explicit knowledge of the end-user [10] [11]. Although it can be reasonably argued that the benefits of modern data driven approach has proven to be extremely beneficial to the society, it remains a serious breach of privacy when the end-user is not aware or has not given explicit consent on using his/her PII data.

There are significant concerns by the end-users on privacy of their data being stored on the cloud. These had been identified by Brill [16], [19]:

- **Risk of Data Breach**: Since the data being stored in the cloud is extensive and relatively concentrated, it makes for a lucrative target for the hackers, and thus, any breach could prove catastrophic [17], [18].
- **"Creepy" Factor**: Consumers are only ever willing to share a certain amount of personal detail with their service providers, and are often at shock when they realize that their service providers know more about them than they intended to.
- **Predictive Policing**: The data that are shared by consumers (knowingly or unknowingly) are often used by government agencies for predictive policing and tracking down potential threats. There have been ample concerns that these could lead to infringement of individual rights.

To address these issues, regulatory bodies around the world have developed privacy policy guidelines to secure data stored in the cloud. However, existing Privacy policy documents are text based and require manual effort to parse and manage. A critical step in automating data privacy management is to make these privacy documents machine process-able so that monitoring tools can interpret the policy rules and metrics defined in them. In our prior work [19], we have developed a semantically rich approach to automate the management of privacy policy documents, using Semantic Web technologies, Natural Language Processing (NLP) and text mining techniques. We created a semantically rich ontology using OWL [19] language to describe the essential components of a privacy policy document and built a database of several privacy policy documents as instances of this ontology. We also developed techniques to automatically extract rules from these policy documents based on deontic logic and demonstrated how it can be used to automate enforcement of data privacy rules.

In this work, we address the challenge of how to automatically track and log every operation performed on PII data that has been shared with providers across the world. The key contribution of this work is the LinkShare system which specifically integrates the blockchain structure with our semantically rich privacy policy ontology to create a secure, trusted, decentralized and auditable Data Privacy Management Framework, comprising of UserBase, PolicyTree and Blockchain. UserBase consists of all legal participants, that is, service providers, users and trusted third parties. For enforcing privacy compliance in access control policies, data protection regulations and relations to enforce access permission based on user choice, form an integral part of the privacy policy ontology contained in PolicyTree. In addition, every data-operation in the form of but not limited to data recording, accessing and sharing, are recorded as part of the blockchain ledger, linked as leaf nodes on the ontology tree. For LinkShare, the block content represents data ownership and viewership permissions shared by members of a private, peer-to-peer network. This read and append only blockchain based ledger is shared between all concerned parties, but access permission relations come from the ontology. This makes sure that the participants can only see appropriate transactions. Moreover, this architecture can also act as a smart contract, since the business terms are embedded in the architecture itself, and every transaction that takes place must follow the set of business rules.

LinkShare systems are by default private and permissioned, as it draws from Semantic Web reasoning concepts, with a strong encryption. It empowers users by enabling them to control all their information and transactions. Being decentralized, LinkShare will not have a central point of failure and thus is better able to withstand malicious attacks. This ensures that the system remains durable and reliable. The handling of transaction processes is done in such a way that the users can trust that transactions will be executed exactly as the privacy protocol commands, removing the need for a trusted third party. Moreover, this disintermediation ensures that two or more parties can carry out an exchange without any oversight, thus strongly reducing or even eliminating counterparty risk.

The data being stored is consistent, timely, accurate, and widely available since LinkShare is based on the principles of Blockchain. Also, we have simplified the Blockchain transaction verification ecosystem and made it faster by utilizing semantic web based reasoning system on the privacy policy ontology. This makes the blockchain transaction times come significantly, instead of having to do costly verification calculations. An only blockchain based system (SmartContract) or only Semantic Web based system would not have worked, since Blockchain does not allow semantic reasoning like an OWL based ontology allows. At the same time, using an OWL based system would have been infeasible, since it does not allow the storage of data reliably and securely. Also, LinkShare facilitates enterprises for the collection, maintenance, use, and disclosure of information, as described in their Privacy Policy including the policy itself, into blockchain. This significantly improves the current multi-party, multi-application solutions and brings it under one, collaboratively managed system.

In this paper, we initially discuss the background and related work in this area. In section III, we describe the architecture of our system. In section IV and V we describe the LinkShare prototype and our validation results. We end with conclusions and future work.

II. RELATED WORK

There exists various solutions to address the need of privacy and access control. The OAuth protocol[5] is one such example, and it is widely used in the industry. But here, the companies themselves act as the trusted, centralized authority. Other such approaches to deal with privacy compliance has typically been role based or attribute based. There also exist ontologies that have been suggested as ways to represent access control concepts, along with legal requirements [1]. But primarily, privacy-preserving methods include differential privacy, a technique that perturbs data or adds noise to the computational process prior to sharing the data, and encryption schemes that allow running computations and queries over encrypted data. Specifically, fully homomorphic encryption (FHE) schemes allow any computation to run over encrypted data, but are currently too inefficient to be widely used in practice.

The National Institute of Standards and Technology (SP 800-144 and SP 800-53) [12] [13] acts as the regulatory guidelines on Security and Privacy in Public Cloud Computing and Security and Privacy in Federal Information

Systems. For identity and Access Management, the NIST standard suggests the usage of SAML(Security Assertion Markup Language) standard. SAML transaction can convey assertions that a user has been authenticated by an identity provider and it can also include information about the user's privileges. Upon receipt of the transaction, the service provider then uses the information to grant the user an appropriate level of access, once the identity and credentials supplied for the user are successfully verified. BitCoins as a system which itself is inherently accountable was invented only a few years back. It uses the concept of blockchain as a publicly verifiable open ledger, and that itself has been proven to be useful in case of trusted computing and auditability. But it mostly has been used as a tool for proof of work computation which is financially lucrative, and as the basis for smart contracts. Chen et al. [15] described a novel framework based on Blockchain with Cloud-based Privacy-aware Role Based Access Control model which may be used for controllability, traceability of data and authorized access to healthcare data resources.

### A. Blockchain

Blockchain technologies have proposed to address the issue of trust and privacy in data security, verifiability and transfer using mathematically designed cryptosystems. This hash-based mathematical protocol allows the system to be cryptographically secure and obfuscated. Per [27], "Blockchain essentially is a distributed database comprising records of transactions or digital events that have been executed and shared among participating parties." Each of these transactions is verified by the consensus of a majority of the participants in the system [22], thus enabling the creation of a distributed consensus in the digital, online world. The characteristics of blockchain technology include features such as smart contracts and smart property. Despite being broadcast between all concerned parties, a blockchain still is implicitly trustable, confidential, secure and auditable platform [9]. However, blockchain protocol does not allow semantically rich policy reasoner to be implemented.

Blockchain is a peer to peer distributed ledger technology [9]. While every member in an ecosystem needs to have its own ledger system and reconcile transaction updates with another member in an expensive and non-standardized operation flows, creating cost efficient business networks are going to be easier with distributed ledger. In addition, because of blockchain's power to trade, manage and service assets in a secure and efficient way, it first gains popularity in financial industry. Blockchains like Bitcoin can get general agreements on both stream of data and computations on the data. An advantage about blockchain is that since we don't need to pay to intermediaries, there is time saving. Another advantage is that blockchains are cheaper more secure and faster than traditional systems. So, blockchains will avoid transactional and legal cost by allowing parties to transact in a secure way.

For our system, we have used the Hyperledger blockchain. As described in [24], "Hyperledger is an open source collaborative effort created to advance cross-industry blockchain technologies" It is an advanced model of blockchain fabric and it is used as a protocol for business to business and business to customer transactions. Record repositories, smart contracts (a decentralized consensus based network, digital assets and cryptographic security) are the central parts of Hyperledger. It allows following the regulations, and when competing businesses work together on the same network, it supports different requirements that they come from. This makes it highly suitable for the current research, since privacy policies are usually of Business to business (B2B) and Business to Customer (B2C) nature.

### B. Semantic Web

In a virtualized service-oriented Big Data scenario, consumers and providers need to be able to exchange information, queries, and requests with some assurance that they share a common meaning. This is critical not only for the data but also for the privacy policies followed by service consumers or providers. While, the handling of heterogeneous policies is usually not present in a closed and/or centralized environment, it is an issue in the open cloud. The interoperability requirement is not just for the data itself, but even for describing services, their service level agreements and their policies for sharing data.

One possible approach to this issue is to employ Semantic Web techniques for modeling and reasoning about services related information. We have used this approach for developing our framework. The Semantic Web deals primarily with data instead of documents. It enables data to be annotated with machine understandable meta-data, allowing the automation of their retrieval and their usage in correct contexts. Semantic Web technologies include languages such as Resource Description Framework (RDF) [25] and Web Ontology Language (OWL) [26] for defining ontologies and describing meta-data using these ontologies as well as tools for reasoning over these descriptions. These technologies can be used to provide common semantics of privacy information and policies enabling all agents who understand basic Semantic Web technologies to communicate and use each other's data and Services effectively.

## III. TECHNICAL APPROACH

The LinkShare system is made of five separate modules, each of which are inter-linked with one another through functionality. The starting point of the system happens through ingestion of privacy policy ontologies. This is handled by a stand-alone module which then passes the PolicyTree created from the ingested ontology to the next module which handles the task of adding or removing relations or processing further updates to the PolicyTree. It is assumed that only a ServiceProvider can start the process of creating the PolicyTree, and add further relations onto it. UserBase can only access and update its own permissions. Upon the execution of any transaction, the Reasoner module will be called, which handles the task of verifying whether the data units used by the current transaction is not in violation of any rules on the PolicyTree pertaining to the concerned user. Based on the result of this Reasoner, a block will get added to the Blockchain. In order to handle queries to the Blockchain, a separate, stand-alone module was made. It takes in *TransactionID* for the current user or ServiceProvider that is initiating the query, and if the User/ServiceProvider is allowed to view/share the results of the Transaction, the result is presented to the party. It is to be noted that LinkShare primarily acts as a distributed ledger with privacy and access control methods.

To define a use-case, let's first go through the main sub-classes of the privacy policy ontology that we have based our implementation on. A snapshot of our version of ontology has been provided in Figure 1. Much of the work in this regard borrows from [19] and expands on it:

- *Collection_Purpose*: This sub-class mainly pertains to the purpose and scope of the data collection. It also allows "collection authority, data transformation actions (such as combining the data with other datasets or performing any data aggregation operations), the duration of the collected data to be stored and managed by the service provider (considering they are the one who will be storing and managing the data) and limitations to the use of the collected data."
- *Data_Protection*: This sub-class mainly pertains to the data storage controls that should be in place. This has been modelled per OWL ontologies for Role based access control [21] and attribute based access control [20].
- *Access_Control*: This sub-class is chiefly acts as an ensuring method to make sure that service providers are using personally identifiable information either as specified in the privacy policy notices or as otherwise permitted by law. "The organization shares PII externally only for the authorized purposes identified in the Privacy Act and/or described in its notice(s) or for a purpose that is compatible with those purposes." Fields that were added to broaden the scope of this class from [19] are IsSharable, IsDataRequested, IsSensitiveData, which acts as a check for all partaking members of the UserBase, end-users and service providers alike.

Consider the following simple scenario: there exists an end-user *U* bound to a service provider *S* based on some set of privacy policies *P*. For a service transaction $T_i$, *U* has to share a subset of personally identifiable information PII with *S*. For this to take place, there must be an existing relation belonging to the subclass Consumer_Consent between the subset PII and Data_Protection. The relation has to establish that the personally identifiable information has been marked with valid Consent_for_Use and Consent_to_share_PII. Also, a relation between subclasses Collection_Purpose->Data_Protection->Access_Control and Personally Identifiable Information should exist. Thus, for that specific transaction $T_i$, if these relations exist and are satisfiable, the transaction passes and is recorded in a sequential blockchain, containing transaction details, as well as fields shared between *S* and *U*. The transaction details are stored as multiple nested blocks with key-value pairs, where the main block key is *TransactionID* and value is the set of PII fields, all being SHA256 encrypted. The leaf blocks for that transaction contains Key-Value pairs of PII field as key and the data shared as the value. For example, consider a user $U_i$ has to share $PII\{i\text{-}j\}$ with $S_i$ for a transaction $T_i$. Before the transaction $T_i$ to pass through, it gets verified against the privacy policy ontology. Say there exists a n-ary relation $R_i$, which is of the form has_<subclass>, such as Personally_Identifiable_Information_<PII Field> -> has_Consent_for_Use, where PII fields are individual personally identifiable information such as "Email", "Address", "PhoneNumber" etc. Once such relation is verified, the next step in the process would be to call the Blockchain module to add the user details shared as Key-Value pair into the blockchain [Figure 2].

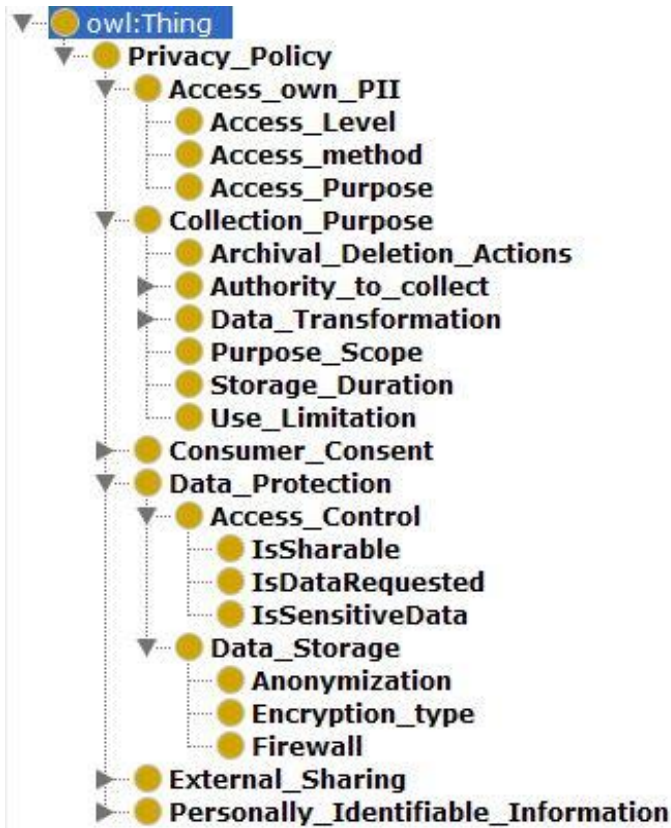

**Figure 1: Ontology Diagram**

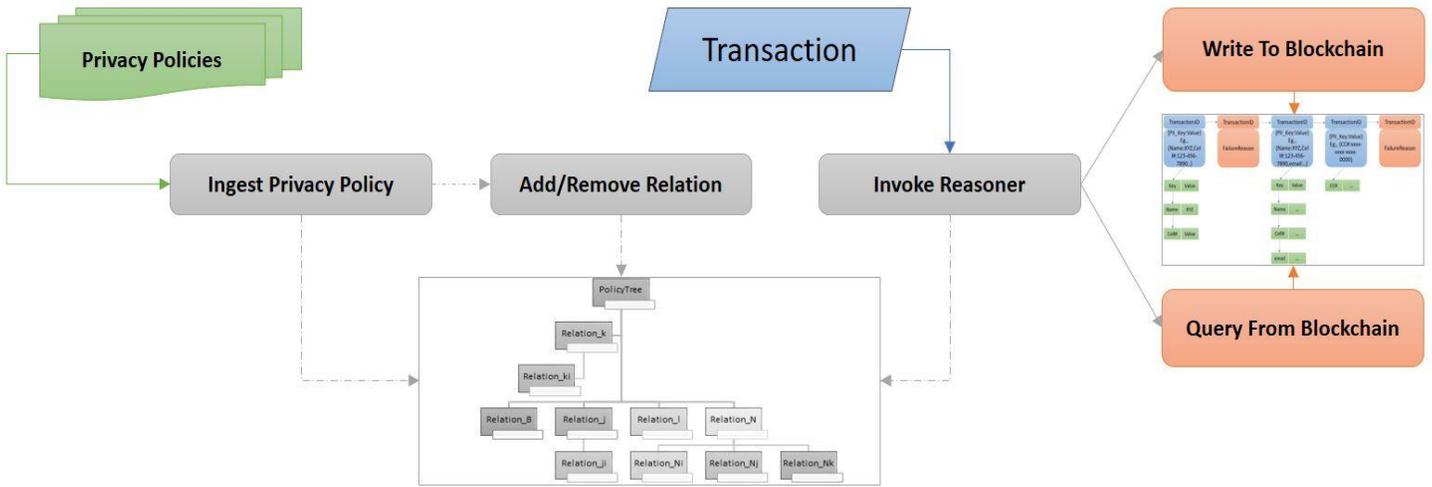

**Figure 2: Process Diagram**

IV. METHODOLOGY

In this paper, we have utilized Semantic Web, NLP and Hyperledger based ChainCode to semi-automate the process of linking and sharing end-user data across businesses. Data is immutable once uploaded. Accordingly, there is no need to decompose tables to reduce redundancy and achieve integrity. We identified key stages in handling the process, and thereby broke it down into 5 interdependent processes in LinkShare [Figure 2], namely:

A. *Ingesting privacy policy*: The privacy policy ontology is first consumed through a module designed to import OWL 2.0 ontologies in the OWL/XML format. It can load an ontology from a local repository, or from Internet. Once loaded, it accesses ontology classes, performs automatic classification of classes and instances of the ontology, creates new instances / individuals and stores them for further manipulation.

B. *Add/Remove relations based on Privacy Policy specifications*: Once the privacy policy has been consumed, this module lets the policy be modified based on the user preference, manipulates ontology classes, instances and properties transparently.

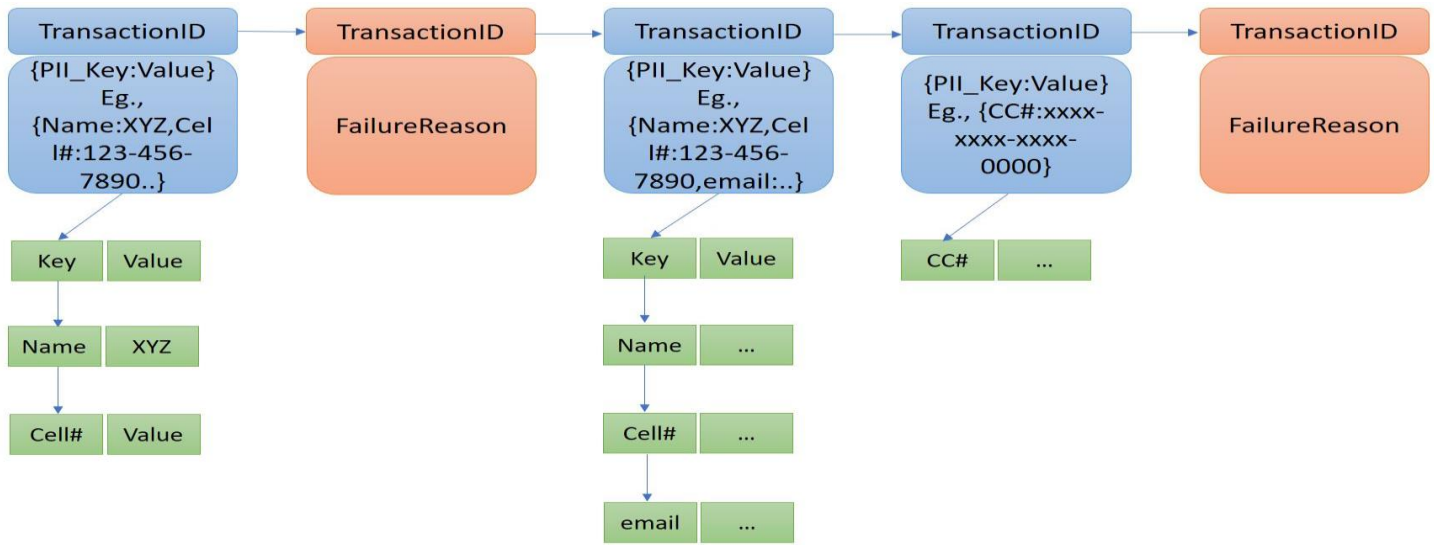

**Figure 3: Visualization of Blockchain Storage**

Also, a new property can be created by sub-classing the Property class, and an existing property can be modified. 'Domain' and 'Range' properties can be specified for the Property as well. A relation is a triple (subject, property, object) where property is a Property class, and subject and object are instances which are subclasses of the 'Domain' and 'Range' defined for the property class. Once the user-preferred relations are created and instances are made, the ontology is passed onto the next step. It is always possible to come back to this module from the next stage in LinkShare.

C. *Invoke Reasoner on every transaction*: Whenever a transaction takes place, the Reasoner is invoked which determines whether the ontology is consistent with the personally identifiable fields required for the transaction and identify subsuming relationships between classes. If the reasoner succeeds, blockchainBranchWrite methods are called with *TransactionID* and {Personally Identifiable Information Field - Value}, otherwise the blockchainWrite is called with *TransactionID* and ReasonerError.

D. *Write to Blockchain*: As illustrated in Figure 3, whenever the blockchain is modified, one of the either two methods are called based on the result of the Reasoner: blockchainBranchWrite or blockchainWrite. blockchainBranchWrite stores the successful transaction as: {*TransactionID* - {key-value}} -> {key - value}, that is, it creates a main block with *TransactionID* as the key, and hashed PII fields with their corresponding values the Value part of the block. The block will always have a branching block which contains PII fields and their corresponding values. The hash of this block can act as the verifiable key for the main block, since any modification in the key-value pair would change the hash of the main block. On the other hand, with the failure of a transaction, blockchainWrite method is called, which stores the failed transaction with *TransactionID* and ReasonerError as a separate block in the blockchain [Figure 3].

E. *Query Blockchain*: To query the blockchain, a separate module for querying the blockchain which accepts a *TransactionID* and fetches the block pertaining to the transaction. Since the blockchain is stored in such a way that any piece of information is directly or indirectly linked with the Transaction it was part of, it is of vital importance that the Query uses *TransactionID* as part of the Query Key term.

V. ARCHITECTURE

The LinkShare system (see Figure 4) consists of 3 distinct entities: UserBase, PolicyTree and BlockchainLedger. UserBase consists of participants of the system, which includes service providers and end-users. Based on the policy ontology tree, a UserBase can either access, share or contribute to the ledger. The access right of an individual component of the UserBase is completely determined by the underlying policy tree. The PolicyTree is created and maintained by the ServiceProvider, subject to change only on maximum consensus between all the concerned parties. Although this might prove to be a hindrance when it comes to implementing further change in the privacy policies, it also provides a mean to make sure that any change is made through public consensus, and are not implemented unilaterally. The PolicyTree and BlockchainLedger forms the backbone of the LinkShare system. The PolicyTree, or ontology, defines all required access control mechanism and privacy policy.

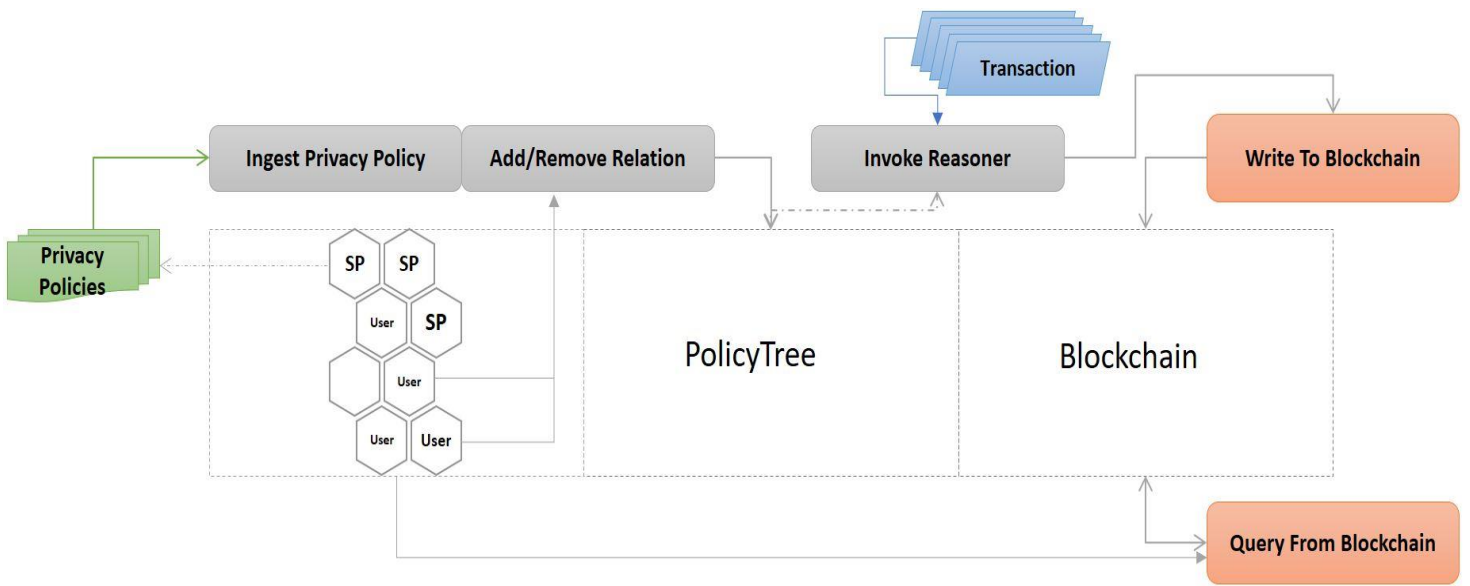

**Figure 4: Architecture Diagram**

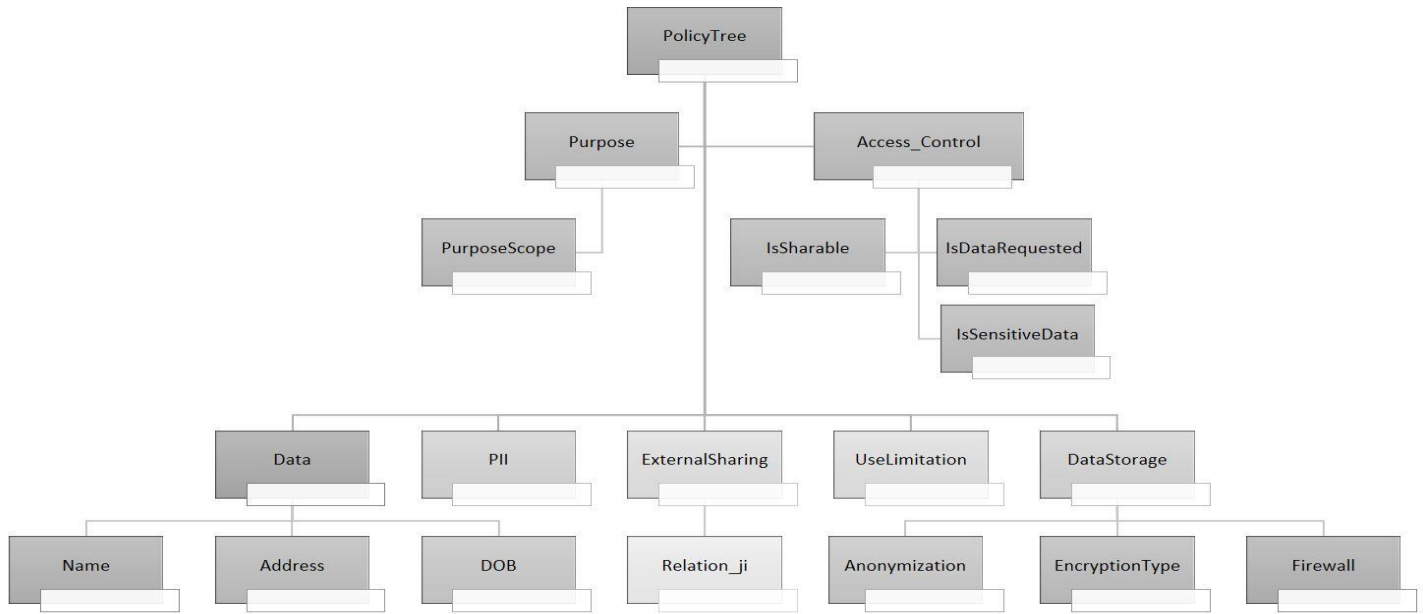

**Figure 5: Visualization of PolicyTree**

The Privacy and access controls are included in the ontology itself. The PolicyTree obligates the service provider to determine and document the legal authority that permits the collection, use, maintenance, and sharing of personally identifiable information (PII), as required by regulatory and compliance bodies. Also, it obligates the end-user and the service provider alike to document purpose(s) for which personally identifiable information (PII) is collected, used, maintained, and shared. The higher level contains legislative requirements for data protection and policies related to operational requirements. The lower levels contain access control relations and policies for handling personal data.

For example, the PolicyTree would contain Data, AccessControl and Purpose as main classes pertaining to how to consume and store data, how to contain access and sharing the data, and how to classify the reason behind accessing, storing or sharing the data. As we can see in Figure 5, drilling down further on Data and AccessControl, we can have granular nodes about individual data points that can be collected, such as Name, Address, DOB, etc., and the relation between an individual user of the system and such individual data points is controlled by AccessControl relation, which specifies whether a node can be accessed based on relations such as IsDataOwner, IsDataController, so on and so forth. The individual data points such as Name, DOB, Address etc. are specific to individual users, and the relations are universal.

Privacy-aware data access policy cannot be easily achieved by traditional access control models. The first reason is that traditional access control models focus on who is performing which action on what data object, privacy policies are concerned with what data object is used for which purpose(s). We propose one purpose-centric access control model. For further clarification into the devised structure, let us consider the following scenario. Say we have service providers Netflix, Amazon, Trusted Third Parties, and an end-user User1. All three will be a part of UserBase. The relations for the individual data points such as Name, Address, DOB etc., such as Is DataOwner, IsDataController, IsDataSharable etc can only be set by the end-user at the time of coming into an agreement with the Service Providers, and these relations can be only be reset at the end-user's behest. Every time any of the concerned parties (service providers, trusted third parties) engages in a transaction, it should be recorded at the end of all the individual data points for that transaction. For example, if Netflix wants to send Amazon Name, ZIP and CreditCard, a new transaction will be recorded, with (Name, ZIP, CreditCard) and (IsSharable, IsDataRequested, IsSensitiveData) requirements to be fulfilled by individual relations between service providers and the individual data points. Every such data point that has been deemed sharable by the end user becomes added as part of a hashed new node at the end of that specific data-point chain, and the transaction is passed.

Security in blockchain-based service computing is a significant backbone of trust-free sharing services. Security is comprised of confidentiality, integrity, and availability; it requires the concurrent existence of: (1) the availability for authorized actions only; (2) confidentiality; and (3) integrity where "improper" means "unauthorized" [14]. As the blockchain is decentralized, the availability of blockchain data does not rely on any third parties. With private and public key cryptography, part of a blockchain's underlying protocol, confidentiality becomes virtually indisputable. Integrity is ensured since the blockchain can be regarded as a distributed file system where participants keep copies of files and agree on changes by consensus.

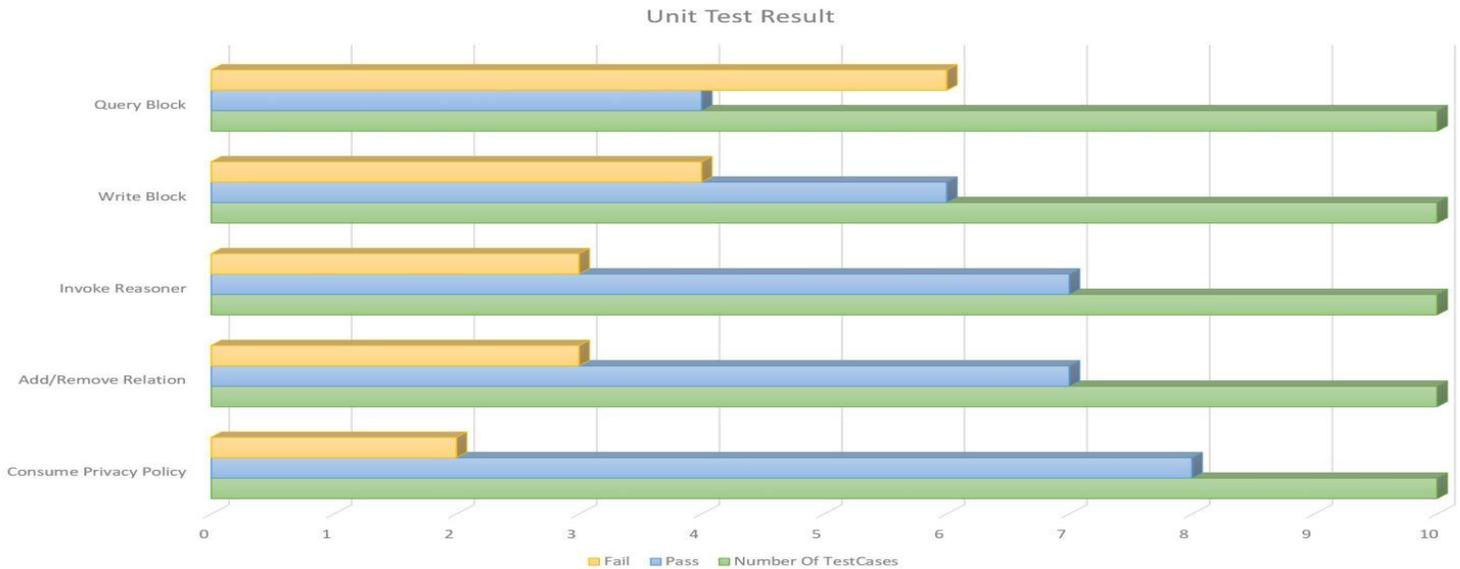

**Figure 6: Visualization of Experimental Result**

It is to be noted that this tree does not contain any data pertaining to the user. The PolicyTree acts as an access control system, and a transaction ledger. This approach not only adds granularity at the service provider to end-user data, it also acts as a verifiable, secure way to comply with any transaction which requires the use of private information. For the end-user, it acts as an assurance that any data that he/she does not want to be shared, won't be shared. From the service provider's point of view, it acts as a verifiable, secure ledger as a mean to facilitate, verify or enforce the privacy policy requirements.

## VI. EXPERIMENTAL RESULTS

To obtain a quantitative performance analysis of the proposed framework, various parameters were measured and evaluated. Small scale scenarios were considered with 10 nodes. In each scenario nodes were split in two sets: Service Providers, i.e., providers of privacy policy resources, resource requesters which execute smart contracts to perform privacy policy-based transactions; End-User, or registered non-institution users in the blockchain. The following parameters were set:

(i) experiment duration: 100 s;
(ii) the Service Providers/End-User ratio: 1:1;
(iii) each Service Providers registered 1 randomly generated End-User;
(iv) each Service Provider sent a randomly generated transaction write request with Key-Value pair and
(v) each End-User sent a new randomly-generated query request every 10 s.

Experiments were performed on a personal computer with Intel i7 4650U CPU at 2.30 GHz, 8 GB of RAM and Windows 10 (64bit) operating system.

In case of consuming privacy policy, pass/fail was determined by how robust the system is in handling xml and owl based privacy policies; Add/Remove relations were run through adding viable and conflicting relations, and hence Reasoner shares similar result. Write to blockchain reflects the number of illegal and legal transaction it handled. Query blockchain shows the actual pass/fail numbers, i.e., out of 10 queries, it could successfully handle 4. ChaincodeInvokeOrQuery invokes or queries the chaincode and if successful, the INVOKE form prints the ProposalResponse to STDOUT, and the QUERY form prints the query result on STDOUT. An absence of standard output or error in connecting to running blockchain was considered as Fail. The results are visualized in Figure 6.

Sharing of personal data is of vital necessity for service-providers in the modern world. Unfortunately, such sharing of data to different third-parties and amongst a single service provider but different modules makes the process awkward and prone to violation of the privacy policy between the end-user and the service provider.

Personal data is extremely vital to the end-user. It is natural to enable end-users to own and control their data without compromising security or limiting the sharing of the service they have opted for. Our architecture enables this by utilizing blockchain platform as purpose-centric access-control model. Based on our architecture, end-users are not required to trust any service-provider or a third-party and are always aware that who is accessing his data and how it will be used. With a decentralized platform and cloud-based central control, making legal and regulatory decisions about collecting, storing and sharing personal data is now simpler. Globally, experimental outcomes show the approach is effective and sustainable for small-to-medium permissioned blockchains.

## VII. FUTURE WORK

In our approach, the privacy policies are consumed as OWL files. After consuming the privacy policy, the policy is stored as linked entities. But expansion and addition of these linked entities as relations are done manually, and thus, it takes extensive skilled manual labour to successfully encode the privacy constraints. Higher scalability can be achieved by properly setting the discovery protocol parameters concerning breadth and width of request propagation as well as response timeout, based on the expected number of participating nodes. Two approaches can be taken as a work-around: firstly, a link finding algorithm such as Path Ranking Algorithm can be utilized to find out new, possible relations that can exist in the policy. These relations can further be presented to the user. Secondly, this overhead in managing the privacy policies by manually adding relations can be mitigated by migrating the privacy policies as constraints of a Smart Contract. For instance, while the Service Provider accesses a personally identifiable information for an end-user, the contract between the concerned parties must include and satisfy the relation "IsDataController".

Adding on to this, any privacy policy can thus be looked upon as a smart contract agreement between B2B and B2C. Being a single client-single service provider system, issues such as batch query and query frequency are handled keeping singular source of query into consideration. We plan to handle query frequency and batch query as part of moving the privacy policy as smart contract. There are several publicly available technologies which handles these issues such as EtherScan and BitCoin Query API. The proposed model can also be equipped with some necessary data management functions which emphasize on further privacy protection: 1. Anonymization: Even before sharing encrypted data, an Access Management module can anonymize data which removes personally identifiable information if necessary. This will come be useful while accessing a set of related data for repurposing (for market basket analysis or other such data manipulations). 2. Communication: Special communication modules can be added to intelligently handle the task of communicating with other related parties for data requests or collaboration. 3. Data backup and recovery: For backing up information on the cloud and to recover information whenever necessary.

In this paper, we have shown a simplified, scaled down version of this system and how it can work. To build upon this, multiple entities can be brought together under the contract and it remains to be researched how this system can be scaled. In the future, we would like to work towards making the privacy policy as a smart contract between B2B and B2C entities, and we are building our framework which would automatically read, understand and transform existing privacy policies in such way.